\documentclass[12pt,preprint]{aastex} 
\usepackage{emulateapj5} 
\usepackage{graphics} 

\newcommand{\lsim}{\raisebox{-.4ex}{$\stackrel{<}{\scriptstyle \sim}$}}

\shorttitle{PO/LCO Atlas of BCDs}
\shortauthors{Gil de Paz et al.}

\begin{document}


\title{Palomar/Las Campanas Imaging Atlas of Blue Compact Dwarf Galaxies: I. Images and Integrated Photometry}


\author{A. Gil de Paz\altaffilmark{1,2,3},
B. F. Madore\altaffilmark{1,3}, and
O. Pevunova\altaffilmark{1}}

\altaffiltext{1} {NASA/IPAC Extragalactic Database, California Institute of Technology, MS 100-22, Pasadena, CA 91125; agpaz, barry, olga@ipac.caltech.edu}
\altaffiltext{2} {Jet Propulsion Laboratory, California Institute of Technology, MS 183-900, Pasadena, CA 91109}
\altaffiltext{3} {The Observatories, Carnegie Institution of Washington, 813 Santa Barbara Street, Pasadena, CA 91101}


\begin{abstract}
We present $B$, $R$, and H$\alpha$ images for a total of 114 nearby
galaxies ($v_{\mathrm{helio}}$$<$4000\,km\,s$^{-1}$) that, with
exception of 9 objects, are classified as Blue Compact Dwarfs
(BCDs). $BR$ integrated magnitudes, H$\alpha$ fluxes and H$\alpha$
equivalent widths for all the objects in the sample are presented. A
new set of quantitative, observational criteria for a galaxy to be
classified as a BCD is proposed. These criteria include a limit on the
$K$-band luminosity (i.e$.$ stellar mass; M$_K$$>$$-$21\,mag), peak
surface brightness ($\mu_{B,\mathrm{peak}}$$<$22\,mag/arcsec$^2$), and
color at the peak surface brightness
($\mu_{B,\mathrm{peak}}$$-$$\mu_{R,\mathrm{peak}}$\lsim1). H$\alpha$
emission is detected in all but three sample galaxies. Typical color,
absolute magnitude, and H$\alpha$ luminosity are
($B-R$)=0.7$\pm$0.3\,mag, M$_B$=$-$16.1$\pm$1.4\,mag, and
log(L$_{\mathrm{H}\alpha}$)=40.0$\pm$0.6 (erg\,s$^{-1}$). Galaxies
morphologically classified as nE and iE BCDs within our sample show
lower H$\alpha$ equivalent widths and redder colors, on average, than
the iI and i0-type BCDs. For most of the galaxies the presence of an
evolved stellar population is required to explain their observed
properties; only the most metal-poor BCDs (e.g$.$ I~Zw~18, Tol~65) are
still compatible with a pure, young burst. The flux-calibrated and
WCS-compliant images in this Atlas are individually available through
the NASA/IPAC Extragalactic Database (NED) image server and
collectively through a dedicated web page at ({\tt
http://nedwww.ipac.caltech.edu/level5/Sept02/Palco\_BCD/frames.html}).
\end{abstract}




\section{Introduction}
\label{intro}

Dwarf galaxies play a crucial role in contemporary theories for the
formation and evolution of galaxies. They are proposed to be the
building blocks from which larger systems have been created by merging
(Kauffmann, White, \& Guiderdoni 1993).


Blue Compact Dwarf galaxies (BCDs) are spectroscopically characterized
by a faint, blue optical continuum accompanied, in most cases, by
intense emission lines. Their current star formation rates (Fanelli et
al$.$ 1988) and neutral gas content (Thuan \& Martin 1981) imply gas
consumption time-scales of $\sim$10$^{9}$\,yr, much shorter than the
age of the Universe. This fact, combined with the low metal abundances
(1/3$>$Z$>$1/50\,Z$_{\odot}$; Hunter and Hoffman 1999), led Searle,
Sargent, \& Bagnuolo (1973) to suggest earlier that either these
objects are young galaxies or that they have experienced an episodic
star-formation history.

The analysis of BCD surface brightness and color profiles (James 1994;
Papaderos et al$.$ 1996a,b; Doublier et al$.$ 1997, 1999; Cair\'{o}s
et al$.$ 2001a,b) combined with {\it HST} color-magnitude diagrams
(Aloisi, Tosi, \& Greggio 1999; Schulte-Ladbeck et al$.$ 1999, 2000,
2001; \"Ostlin 2000; Crone et al$.$ 2000, 2002; Drozdovsky et al$.$
2001) has shown the existence of an underlying stellar population at
least a few 10$^{9}$\,yr old in most of these galaxies, i.e. these are
not purely ``young'' galaxies. In this sense, Legrand et al$.$ (2000)
and Crone et al$.$ (2000) have recently suggested that the formation
of this evolved stellar population could have taken place at a low but
continuous rate rather than in a purely bursting mode (see also Heller
et al$.$ 2000 concerning the lack of other evidence for random star
formation in BCDs). Only in the case of the most metal poor BCDs
(Z$<$1/20\,Z$_{\odot}$; Izotov \& Thuan 1999) are the results
regarding their evolutionary status still contradictory, mainly due to
uncertain distances (\"Ostlin 2000) and contamination of the galaxy
outer halo colors by nebular emission (Izotov et al$.$ 2001a,b;
Papaderos et al$.$ 1998, 2002).

Understanding of the role played by the collective supernovae-driven
winds in the mass (i.e. luminosity) and chemical evolution, and in the
propagation of the star formation in these galaxies is also mandatory.
In this sense, the evolution of these winds in low-mass galaxies like
the BCDs is thought to lead to the loss of a (still highly uncertain)
fraction of metals (De Young \& Heckman 1994; Mac-Low \& Ferrara 1999;
Silich \& Tenorio-Tagle 2001), making these galaxies arguably one of
the major polluters of the intergalactic medium (Ferrara \& Tolstoy
2000).

In order to shed light on some of these questions we have obtained
$BR$H$\alpha$ imaging data for a statistically significant sample of
BCD galaxies. This sample includes 114 galaxies, 105 of them
classified as BCDs. Similar recent studies include Papaderos et al$.$
(1996a,b), Telles \& Terlevich (1997), Doublier et al$.$ (1997, 1999),
and Cair\'{o}s et al$.$ (2001a,b). However, these studies analyze
samples that contain at least a factor of 4 fewer galaxies than our
sample and, excepting the recent work of Cair\'{o}s et al$.$ (2001b),
none of these studies include emission-line imaging data.  \\

In Section~\ref{sample} we describe the properties of our sample. A
new set of observational, quantitative criteria to define the Blue
Compact Dwarf class of galaxies is proposed in
Section~\ref{newclass}. The observations and reduction procedures are
described in Sections~\ref{observations} and
\ref{reduction}. Sections~\ref{maps} and \ref{images} present the
images of the galaxies along with their morphological
classification. The results from the analysis of the integrated
photometry of the sample are given in Section~\ref{analysis}. We
summarize our conclusions in Section~\ref{conclusions}. In Paper II we
will analyze the structural properties of the sample, both in the
continuum and in the emission-lines, by carrying out a decomposition of
the surface brightness profiles, and presenting the concentration and
asymmetry parameters in the different bands.

\section{Sample}
\label{sample}

The galaxies in our sample have been selected from several exploratory
studies including the Markarian lists (Mrk -- Markarian, Lipovetskii,
\& Stepanian 1981), Second Byurakan (SBS -- Markarian, Stepanian, \&
Erastova 1986), University of Michigan (UM -- MacAlpine \& Williams
1981), Palomar Objective-prism X (POX -- Kunth, Sargent, \& Kowal
1981), Hamburg/SAO (HS -- Ugryumov et al$.$ 1999), Universidad
Complutense de Madrid (UCM -- Zamorano et al$.$ 1994, 1996; Alonso et
al$.$ 1999), and Tololo (Tol -- Smith, Aguirre, \& Zemelan 1976)
surveys, as well as the lists of Zwicky \& Zwicky (1971), and Haro
(1956). Mrk and SBS galaxies were originally selected by their
ultraviolet excess on photographic plates taken with the
objective-prism tecnique. POX, HS, UM, UCM, \& Tol galaxies were
discovered by the presence of emission lines, [OIII]$\lambda$5007\AA\
(UM, HS, \& Tol), H$\alpha$ (UCM), or both (POX), in objective-prism
spectra. Finally, galaxies in the lists of Zwicky \& Zwicky (1971) and
Haro (1956) were selected according to their compactness and blue
colors as inferred from photographic plates taken through 3 different
filters (blue, red, \& infrared).

From the list of BCD galaxies discovered in these surveys we have
selected objects with heliocentric recession velocities lower than
4000\,km\,s$^{-1}$. This limit was set by the longest-wavelength
H$\alpha$ filter (PO 6640) available within the Palomar Observatory
narrow-band filters set. Our sample also includes the six Virgo
Cluster BCDs analyzed by Popescu et al$.$ (2002) and the highly
obscured Local Group galaxy IC~10. Although this latter galaxy has
been ordinarily classified as a dwarf irregular, the recent study of
Richer et al$.$ (2001) shows that its structural properties resemble
more those of BCD galaxies (see Papaderos et al$.$ 1996a). Our final
sample consists of 114 galaxies.

It is worth mentioning that this sample does not include all the
galaxies classified as BCDs in those surveys (noteworthy missing
objects are UM~461, UM~462, or UM~465). Our sample cannot either be
considered a complete sample since it has been selected from surveys
that used different techniques and, therefore, are affected by
different selection criteria.

In Table~\ref{table1} we summarize the basic properties of the
galaxies in this Atlas. In the few cases where the distance to the
galaxy had been determined by measuring the magnitude of the tip of
the RGB we adopted the distance value given by this method (see Lee,
Freedman, \& Madore 1993). In the vast majority of the cases we
computed the distance using the Galactic Standard of Rest velocity of
the galaxies assuming a Hubble constant of
70\,km\,s$^{-1}$\,Mpc$^{-1}$ (Freedman et al$.$ 2001). Distances to
the galaxies in the Virgo Cluster were adopted to be 16\,Mpc (Macri et
al$.$ 1999). For 49 galaxies with published data the
[NII]$\lambda$6584\AA/H$\alpha$ line ratios required to correct the
H$\alpha$ fluxes measured from the [NII]$\lambda\lambda$6548,6584\AA\
doublet contamination were taken from Guseva, Izotov, \& Thuan (2000),
Izotov \& Thuan (1999), Terlevich et al$.$ (1991), Kennicutt (1992),
V\'{\i}lchez (1995), Masegosa, Moles, \& Campos-Aguilar (1994),
Kobulnicky \& Skillman (1998), Kniazev et al$.$ (201), van Zee et
al$.$ (1998a), Hunter \& Hoffman (1999), Steel et al$.$ (1996), Vacca
\& Conti (1992), Thuan, Izotov, \& Lipovetsky (1995), Popescu \& Hopp
(2000), Garnett (1990), Thuan, Izotov, \& Foltz (1999), Jansen et
al$.$ (2000), Rego et al$.$ (1998), and Augarde et al$.$ (1990). In
those cases where the [NII]$\lambda$6584\AA/H$\alpha$ line ratio was
not available we adopted an average value of 0.055 computed using the
49 galaxies in our sample that do have [NII]$\lambda$6584\AA/H$\alpha$
data. For comparison the average value obtained by Hunter \& Hoffman
(1999) for the BCDs in their sample was 0.091.

Equatorial coordinates, Galactic extinction (A$_B$) and heliocentric
radial velocity values for the sample were taken from NED (see
Table~\ref{table1}). Galactic extinction data are derived from the
100\,$\mu$m maps of Schlegel, Finkbeiner, \& Davis (1998). The values
obtained are typically smaller than A$_B$=0.5\,mag, except for IC~10
(6.2\,mag), II~Zw~40 (3.5\,mag), and NGC~2915 (1.2\,mag). In the case
of IC~10 the very low Galactic latitude makes the foreground
extinction correction very uncertain.

\section{Definition of Blue Compact Dwarf galaxy}
\label{newclass}
In the past many different criteria have been used to classify a
galaxy as a BCD or not. These criteria were commonly based on the
galaxy's luminosity and its morphological properties (Zwicky \& Zwicky
1971; Thuan \& Martin 1981) although definitions based on their
spectroscopic properties are also found in the literature (Gallego et
al$.$ 1996). Moreover, galaxies morphologically classified as BCDs are
sometimes confused with spectroscopically-classified objects, such as
``isolated extragalactic HII regions'' (Sargent \& Searle 1970), ``HII
galaxies'' (Terlevich et al$.$ 1991) or ``Sargent-Searle objects''
(SS; Salzer, MacAlpine, \& Boroson 1989). For instance, although the
``isolated extragalactic HII regions'' of Sargent \& Searle (1970) and
SS objects of Salzer et al$.$ (1989) can be undoubtedly classified as
BCDs, many HII galaxies in Terlevich et al$.$ (1991) are significantly
brighter than a BCD.

The original definition of ``Compact Galaxy'' comes from Zwicky (1970)
where he defined as ``compact'' any galaxy (or any part of a galaxy)
whose surface brightness is brighter than 20\,mag/arcsec$^2$ in any
chosen wavelength range. The term ``blue'', as used by Zwicky, refers
to those galaxies satisfying the previous condition on both blue and
red plates (Zwicky \& Zwicky 1971). Later on, Thuan \& Martin (1981)
introduced the term ``Blue Compact Dwarf'' referring to those galaxies
having absolute blue magnitudes fainter than M$_B$=$-$18.15\,mag
(H$_0$=70\,km\,s$^{-1}$\,Mpc$^{-1}$), diameters less than 1\,kpc, and
strong emission-lines superposed on a blue continuum. More recently,
Gallego et al$.$ (1996) spectroscopically classified as BCDs those
galaxies showing intense, high-excitation emission lines and low
H$\alpha$ luminosity
(L$_{\mathrm{H}\alpha}$$<$10$^{41}$\,erg\,s$^{-1}$; for
H$_0$=70\,km\,s$^{-1}$\,Mpc$^{-1}$). Finally, some variations on these
definitions can be also found in the literature for selecting samples
of BCD galaxies (Doublier et al$.$ 1997; Kong \& Cheng 2002).


In this paper we attempt to unify the concept of BCD by putting
forward a new set of quantitative classification criteria. Using these
criteria we are able to better include within the BCD class galaxies
sharing common physical properties and evolutionary status and
segregate them from other types of objects like dwarf irregulars
(dIrr) and dwarf ellipticals (dE) or more massive star-forming
galaxies. These criteria are sufficiently inclusive so as to also
recover most of the galaxies traditionally classified as BCD.\\


\noindent {\bf Blue}. Probably the most ill-defined property of the
``Blue Compact Dwarf galaxies'' is the color. Although the presence of
a blue continuum in the optical spectra was already required by Thuan
\& Martin (1981) in their definition of BCDs, it was only a
qualitative criterion. The observational criterion traditionally used
has been the color of the highest surface brightness component since
it was generally the only component detected in surveys using
photographic plates. Thus, in order to establish a more quantitative
criteria in this same sense we have determined the peak surface
brightness (PSB) and the color at this peak surface-brightness from
the surface-brightness profiles (Paper II) of the galaxies in our
sample, both corrected for Galactic extinction. In order to reduce the
effects of different seeing between the images we have averaged the
color within the inner 3\,arcsec of the profile; in those galaxies
with optical diameter $\leq$20\,arcsec we averaged the inner
1\,arcsec.


Figure~\ref{figure1}a shows the distribution in color at the PSB
measured from the galaxies' surface-brightness. We have also plotted
the colors at the PSB obtained for a sample of dIrr (Parodi, Barazza,
\& Binggeli 2002) and dE (Jerjen, Binggeli, \& Freeman 2000) with
published surface-brightness and ($B-R$) color profiles. This figure
shows a clear segregation between the color at the PSB of galaxies
previous classified as BCDs and that of dIrr and dE, although an
overlap is present. We have decided to impose a limit of ($B-R$) color
at the peak of
$\mu_{B,\mathrm{peak}}$$-$$\mu_{R,\mathrm{peak}}$\lsim1. Using only
this criterion many dE and most of the dIrr galaxies would be
classified as BCDs. However, the combination with other criteria will
improve the situation significantly (see below). Even with this wide
limit some galaxies traditionally classified as BCD by other authors
are apparently quite red. Particularly noticeable are the cases of
UCM~0049$-$0045, UCM~1446$+$2312, and VCC~0001, that show peak color
redder than $\mu_{B,\mathrm{peak}}$$-$$\mu_{R,\mathrm{peak}}$=1.2 and
should not be classified as BCDs.\\

\noindent {\bf Compact}. With regard their compactness, Thuan \&
Martin (1981) set a upper limit to the optical diameter of BCDs of
1\,kpc. However, observations carried out with CCDs during the 90's
have shown the presence of a very extended (up to a few kpc)
low-surface-brightness component in many objects that were previously
classified as BCDs. Therefore, the term ``compact'' in BCDs should be
related more with the size of the high-surface-brightness component
than with the total optical size. In other words, the compactness
criterion should be more a surface brightness limit, like that used in
Zwicky (1970), than a real physical-size limit.


In Figure~\ref{figure1}b we show the distribution of peak surface
brightness (PSB) measured from the galaxies $B$-band surface
brightness profiles (to be presented in Paper II). This figure shows
the PSB of BCDs being significantly brighter than that of dIrr (Parodi
et al$.$ 2002; Patterson \& Thuan 1996) and dE (Jerjen et al$.$
2000). A fairly good separation between the different types is
achieved imposing a limit of
$\mu_{B,\mathrm{peak}}$$<$22\,mag/arcsec$^2$ to the PSB of BCDs. A
total of 9 galaxies in our sample show PSB fainter than this value, so
they should not be classified as BCDs. However, an individualized
analysis of these objects show that 6 of them are $\leq$20\,arcsec in
diameter that makes the atmospheric seeing to significantly dim their
PSB. Five of these objects (HS~0822$+$3542, UM~382, UM~417,
SBS~0940+544C, NGC~4861) also show cometary morphology. Since the PSB
have been determined from the azimuthally-averaged surface-brightness
profiles, the PSB in these cases is measuring surface brightness at
the peak of the low-surface-brightness component instead that the
value at the burst located at the edge of the galaxy (Paper
II). Finally, two of these objects should certainly not be classified
as BCDs but as dIrr, II~Zw33~B (Walter et al$.$ 1997) and UGC~4483
(van Zee, Skillman, \& Salzer 1998b).

The physical reason for the segregation observed in $B$-band PSB
between the different galaxy types is the presence in the BCDs of a
recent star formation event that outshines the low-surface-brightness
component. This event may be accompanied by blue optical colors and
strong emission-lines (as in objects spectroscopically classified as
HII galaxies). In dIrr and dE the recent star formation (if present)
is comparatively less active and the PSB is consequently fainter and
redder.


It is important to note that the fact that the segregation between
BCDs and other types of dwarfs by PSB is better than the segregation
by the color at the PSB is partially due to the different contribution
of the red supergiants to the luminosity and color evolution of the
burst (Doublier et al$.$ 2001a).\\

\noindent {\bf Dwarf}. One of the most important physical parameters
driving the evolution of galaxies is the mass (see Brinchmann \& Ellis
2000 and references therein). This is particularly important in the
case of low-mass galaxies like the BCDs where it controls the
formation of density waves or not. In this sense, the $B$-band
luminosity cutoff imposed by Thuan \& Martin (1981) was thought as a
limit in the stellar mass of BCDs. However, the $B$-band luminosity is
a very poor tracer of the stellar mass in a galaxy. Here we propose to
use the $K$-band luminosity as a more reliable measure of the stellar
mass in these galaxies (Gil de Paz et al$.$ 2000a; P\'erez-Gonz\'alez
et al$.$ 2003a,b). In this regard it is worth mentioning that the
assumption of different, plausible star formation histories within a
galaxy may lead to changes in the mass-to-light ratio as high as a
factor or 7 in the $B$-band but only a factor of 2 in $K$ (Bell \& de
Jong 2001).


Unfortunately, the number of studies of BCDs in the near-infrared is
still small and they are limited to a very few objects each (James
1994; Vanzi, Hunt, \& Thuan 2002; Doublier, Caulet, \& Comte 2001b;
Noeske et al$.$ 2003, submitted). In Figure~\ref{figure1}c we have
plotted the ($B-K$) and ($B-R$) colors for the 21 galaxies in our
sample included in the 2MASS Second Incremental Release Extended
Source Catalog (Jarrett et al$.$ 2000). Dotted lines in this diagram
represent the model predictions for a composite stellar population
formed by a Z$_{\odot}$/5 metal-abundance burst with burst strength 1\%
(in mass) and age between 3.5 and 10\,Myr. Thin, solid-lines
correspond to the time evolution predicted by the models for the same
burst with strength between 100\% and 0.01\%. The models used here are
those developed by Gil de Paz et al$.$ (2000a) and Perez-Gonz\'alez et
al$.$ (2003a,b) which are based on the evolutionary synthesis models
of Bruzual \& Charlot (2003, unpublished). We have assumed that 15\%
of the Lyman continuum photons escape from the galaxy (or are absorbed
by dust) before the ionization of the surrounding gas (see e.g$.$ Gil
de Paz et al$.$ 2000a). This figure shows that there is a very good
linear correlation between the integrated ($B-R$) and ($B-K$) color of
these objects, which is also expected from the predictions of the
models for burst strength values lower than 1\% or ($B-R$) colors
redder than $\sim$0.5. The best fit to this correlation is
\begin{equation} 
(B-K) = 1.361 \times (B-R) + 1.63 \ \ \ ;\ \ \sigma = 0.25\,\mathrm{mag}
\label{bkvsbr}
\end{equation} 
As we commented above this relation is applicable only when the
($B-R$) color is $>$0.5\,mag. However, as we will show in
Section~\ref{analysis} virtually all the BCD galaxies with ($B-R$)
color bluer than 0.5\,mag have $B$-band absolute magnitude fainter
than $M_B$=$-$16.5, so they can undoubtedly classified as BCDs.


The average ($B-K$) color of the 21 sample galaxies within 2MASS is
2.82$\pm$0.42. If we now apply this average ($B-K$) color to the limit
in $B$-band luminosity imposed by Thuan \& Martin (1981) we end up
with an equivalent limit in $K$-band luminosity of $M_K$$>$$-$21\,mag
(for H$_0$=70\,km\,s$^{-1}$\,Mpc$^{-1}$). In Figure~\ref{figure1}d we
show the frequency histogram of $K$-band absolute magnitudes of the
galaxies in our sample obtained applying the
Equation~\ref{bkvsbr}. Noteworthy, from the 9 galaxies in our sample
showing $M_B$$<$$-$18.15\,mag only 5 are brighter than
$M_K$=$-$21\,mag. Within them IC~10 is probably above this limit
because of its uncertain galactic extinction correction. The other 4
galaxies should not be classified as BCD galaxies, II~Zw~33, Mrk~7,
Tol~1924$-$416, and Mrk~314. The galaxies showing
$M_B$$<$$-$18.15\,mag but $M_K$$>$$-$21\,mag (KUG~0207$-$016A,
Mrk~400, Haro~2, Pox~4) are probably relatively low-mass objects
experiencing a very massive burst that makes their integrated colors
bluer than the average. Individualized near-infrared observations are
necessary to confirm this.\\


Summarizing, we propose that, in order to be classified as a BCD, a
galaxy has to fulfill the following observational, quantitative
criteria: (1) it has to be {\it blue},
$\mu_{B,\mathrm{peak}}$$-$$\mu_{R,\mathrm{peak}}$\lsim1, (2) {\it
compact}, $\mu_{B,peak}$$<$22\,mag/arcsec$^2$, and (3) {\it dwarf},
$M_K$$>$$-$21\,mag. As we have shown above these criteria recover most
of the galaxies traditionally classified as BCDs and also allow to
segregate the BCDs from other types of dwarf galaxies like dIrr or dE.

\section{Observations}
\label{observations}

We have observed the whole sample of 114 galaxies in $B$, $R$, and in
the light of H$\alpha$ during a total of 11 observing runs between
June 2001 and July 2002. Nine of the observing runs were carried out
at the Palomar Observatory 60-inch telescope using CCD Camera with the
2048$\times$2048 pixel CCD\#13 attached. The pixel scale was 0.378
arcsec/pixel. Two additional observing runs were carried out at the du
Pont 100-inch telescope in Las Campanas Observatory (Chile) between
February 8-9 2002 and March 5-9 2002. We used the Direct CCD with the
2048$\times$2048 pixel Tek5 CCD attached, which, placed at the
Cassegrain focus of the telescope, gives a scale of 0.260
arcsec/pixel. In Table~\ref{table2} we give a summary of the observing
runs and the properties of the detectors used (gain, readout noise,
etc).

Typical exposure times in $B$, $R$, and H$\alpha$ were, respectively,
15, 45, and 90\,min at Palomar Observatory and 15, 15, and 30\,min at
Las Campanas. A total of 86 galaxies were observed at Palomar
Observatory and 28 at Las Campanas. The seeing (FWHM) in the images
ranged from 0.6\,arcsec (Tol~002 in $R$) to 4.3\,arcsec (Haro~14
in $B$). The median seeing values in $B$, $R$, and H$\alpha$ were 1.8,
1.6, and 1.5\,arcsec, with 80 per cent of the galaxies having seeing
better than 2.5, 2.0, and 1.9\,arcsec, respectively. The image quality
at Las Campanas Observatory was significantly better than at Palomar
Observatory. In this sense, the median seeing of the $B$, $R$, and
H$\alpha$ images taken at Las Campanas were 1.1, 0.9, and 0.9\,arcsec,
respectively, whereas at Palomar Observatory the median seeing values
achieved were 2.0, 1.7, and 1.6\,arcsec. In Table~\ref{table3} we give
a summary of the characteristics of some of the images contained in
this Atlas, including observatory, telescope, date, exposure time,
number of frames, airmass, and FWHM (the complete table is available
in the electronic version of the paper). The date of observation given
in this table corresponds to the civil date at the start of the
observing night.

In Figure~\ref{figure2} we give the spectral response functions of the
filters used in this Atlas as provided by the manufacturers. Thin
continuous-lines correspond to the filters used at the Palomar
Observatory 60-inch telescope and broad-lines to those used at Las
Campanas 100-inch telescope. This figure clearly shows that the
Johnson-$B$ and Cousins-$R$ filters used in this Atlas are very
similar to those used by Landolt (1992a) to build his list of standard
stars (dotted-line) and also similar to those originally used to
define the Johnson-Cousins system (dashed-line; $B$-band: Azusienis \&
Straizys 1969; $R$-band: Bessell 1990). It is worth noting that our
filters as plotted have been convolved with the spectral response
functions of the corresponding detectors. This could be, in part, the
reason for the apparently-poorer spectral response of our filters to
wavelengths longer than $\sim$6500\,\AA\ compared to those used by
Landolt (1992a) and Bessell (1990). Narrow-band filters used at
Palomar and Las Campanas were 20\,\AA\ and 65\,\AA\ wide,
respectively. The use of narrower H$\alpha$ filters at Palomar
Observatory reduced the contribution of the continuum to the photon
noise of the H$\alpha$ images, partially compensating for both the
poorer image quality of the H$\alpha$ images obtained at Palomar
Observatory and the smaller collecting area of the 60-inch telescope.

\section{Reduction}
\label{reduction}

The images were reduced following the standard procedures of bias
subtraction and flat-fielding using the IRAF\footnote{IRAF is
distributed by the National Optical Astronomy Observatories, which are
operated by the Association of Universities for Research in Astronomy,
Inc., under cooperative agreement with the National Science
Foundation.} task {\sc ccdproc}. The dark current in both detectors
used was negligible ($\leq$1\,count/hour/pixel) and no dark
subtraction was applied. In those cases where three or more frames in
one band were obtained for the same object the cosmic rays were
rejected using the IRAF task {\sc imcombine}. When only one or two
images were available we removed the cosmic rays interactively using
the IRAF tasks {\sc cosmicrays} and {\sc credit} within the CRUTIL
package.

\subsection{Flux-calibration}
\label{calibration}

Considering the relatively small number of nights that were
photometric (12 out of 49) we decided to use the following strategy in
order to flux-calibrate our images. During the non-photometric nights
at Palomar and Las Campanas observatories we obtained very deep
exposures in $R$ and H$\alpha$. During the photometric nights the
adopted follow-up strategy was slightly different at each site. At
Palomar we took short (600-900\,s) $B$ and $R$ exposures of the same
objects already observed in $R$ and H$\alpha$, along with images of a
large number of photometric standards (Landolt 1992a). Then, using the
fluxes of the field stars in the calibrated $R$-band, deep $R$ and
H$\alpha$ images along with the filter+detector response functions
(see Figure~\ref{figure2}) we cross-calibrated our deep $R$ and
H$\alpha$ images (see Appendix~\ref{appendixA}). At Las Campanas,
during the photometric nights we observed both objects and
spectrophotometric standards (Hamuy et al$.$ 1992; Landolt 1992b)
through all $B$, $R$ and H$\alpha$ filters. During these nights we
also took short $B$ and $R$ exposures of the objects already observed
in $R$ and H$\alpha$ during previous non-photometric nights. The flux
calibration of the H$\alpha$ images taken during non-photometric
conditions was carried out in the same manner as the Palomar case (see
Appendix~\ref{appendixA}). The coefficients obtained from the
calibration are given in Table~\ref{table4}. This strategy has allowed
us to make optimal use of telescope time, achieving relatively small
photometric errors (see Table~\ref{table5}). In order to check the
reliability of the H$\alpha$ calibration based on the use of $R$-band
calibrated data, we compared the results obtained from this method
with those resulting from the observation of spectrophotometry
standards through the H$\alpha$ filters during the Las Campanas
photometric nights. The calibration results agreed to within 5\%.

The continuum-subtracted H$\alpha$ images of our sample were obtained
from the $R$-band and line+continuum H$\alpha$ images making use of
the equations given in Appendix~\ref{appendixA}. Briefly, we compared
the flux in counts of a large number of field stars ($>$30) both in
the $R$-band and H$\alpha$ images. Then, using the IRAF task {\sc gauss} the
image with the best seeing was convolved with a gaussian kernel to
match the seeing of the worst-quality image. Finally, we divided the
$R$-band image by the corresponding scaling factor and subtracted it
from the line+continuum H$\alpha$ image. Residuals in the
continuum-subtracted images due to the presence of very bright field
stars were removed interactively using the IRAF task {\sc credit} within the
CRUTIL package. For more details about this procedure and the
flux-calibration of the resulting continuum-subtracted H$\alpha$ image
the reader is refereed to Appendix~\ref{appendixA}.

\subsection{Astrometry}
\label{astrometry} 

In order to perform the astrometry on the images we first aligned all
the frames in the same band taken on the same night for the same
object using the IRAF task {\sc imshift} and the positions of a few
field stars. Then, using the positions of a larger number of field
stars ($>$100), as given by the IRAF task {\sc starfind}, we
registered the images in the different bands using the {\sc geomap}
and {\sc geotran} tasks. Finally, once all the frames were combined
and the various bands registered, we used the IRAF tasks {\sc
starfind} and {\sc ccxymatch} to cross-correlate the positions of the
field stars in pixels with their FK5 equatorial coordinates in the
USNO-A2.0 catalogue. The cross-correlation was done using the
positions of the stars in our final $R$-band image. The H$\alpha$
(line+continuum) image was used in those cases where a large number of
USNO-A2.0 stars were saturated in the $R$-band image (e.g$.$ in those
cases of exceptional seeing). Once the plate solution was computed and
stored in the image header (using the IRAF tasks {\sc ccmap} and {\sc
ccsetwcs}) we copied the WCS information to the rest of the bands
using the task {\sc wcscopy}. Typical (rms) errors in the astrometric
calibration of our images are 0.1 arcsec both in RA and DEC.

\section{Maps and Morphological Classification}
\label{maps}

In Figure~\ref{figure3} we present greyscale maps corresponding to the
$R$-band and to continuum-subtracted H$\alpha$ images obtained for the
114 galaxies in our sample. A blow-up of the nuclear region is also
shown in the upper right-hand corner of each image, rescaled in
intensity to emphasize details. The size of the compass shown in each
of the images is 0.5\,kpc at the distance of the corresponding
galaxy. The average limiting surface brightness of the main images
(3$\sigma$) are 24.6\,mag\,arcsec$^{-2}$, 24.3\,mag\,arcsec$^{-2}$,
and 7$\times$10$^{-17}$\,erg\,s$^{-1}$\,cm$^{-2}$\,arcsec$^{-2}$ for
$B$, $R$, and H$\alpha$, respectively.

Within the galaxy sample studied only three galaxies were not detected
by us in H$\alpha$ emission, namely Mrk~709, UCM~1446+2312, and
UGC~4703~NOTES02. In the case of Mrk~709 the galaxy recession velocity
provided by NED (1,197\,km\,s$^{-1}$), which was taken from the
21cm-line velocity data published by van Zee, Haynes, \& Giovanelli
(1995), differs significantly from that published by Terlevich et
al$.$ (1991) ($\sim$15,000\,km\,s$^{-1}$) and recently confirmed by
E$.$ P\'{e}rez-Montero (private communication). This leads us to
suggest that the low-redshift HI emission detected by van Zee et al$.$
(1995) is not related to Mrk~709, but maybe with another object in the
beam that, according to our data, does not show significant ionized
gas emission. For UCM~1446+2312 the uncertainty in the recession
velocity (determined from the position of the H$\alpha$ line in
objective-prism photographic plates; Alonso et al$.$ 1999) is
$\sim$$\pm$1,200\,km\,s$^{-1}$. Therefore, the line-emission arising
from this object might well be beyond the wavelength region covered by
the 20\,\AA-wide filter used. Finally, UGC~4703~NOTES02 forms with
UGC~4703~NOTES01 an interacting pair where only the latter shows
H$\alpha$ emission. This suggests that UGC~4703~NOTES02 may have
either exhausted most of its gas early during its evolution, or it may
have lost it as consequence of the interaction with
UGC~4703~NOTES01. Is it important to note that, as already pointed out
by Cair\'{o}s et al$.$ (2001a) and we have discussed in
Section~\ref{newclass}, most of the BCD galaxies were originally
identified by their high surface brightness and blue optical
colors. Therefore, some of them may be expected to show very low
H$\alpha$ equivalent widths (EW) (i.e$.$ they are BCDs but they should
not be then classified as HII galaxies). Some examples of this may be
Mrk~1423 and I~Zw~115 (see Section~\ref{analysis}).

The maps presented in Figure~\ref{figure3} show that most of the
galaxies in our sample have an extended, low-surface-brightness halo
visible in the continuum images. Such halos, commonly associated with
an evolved stellar population (Papaderos et al$.$ 1996; Cair\'{o}s et
al$.$ 2001a,b and references therein), are usually more extended than
the region showing H$\alpha$ emission, except perhaps in the case of
the most compact objects (I~Zw~18, HS~0822+3542, Pox~186). In Paper II
we analyze the structural properties of this low-surface brightness
component. With regard to the morphology in H$\alpha$ the inspection
of our images show that 83 galaxies in our sample (73\%) show star
formation activity distributed over more than one region. In
particular, in 45 galaxies (39\%) we are able to distinguish 4 regions
or more. Only 28 galaxies show their morphology in the light of
H$\alpha$ to be dominated by only a single, nuclear HII region. It is
worth noting that the median distance and seeing for those galaxies
showing only one single star-forming region are 23\,Mpc and
1.4\,arcsec respectively, while the galaxies with multiple regions are
typically at 19\,Mpc and have 1.5\,arcsec seeing. Clearly, the
difference between these values is not large enough to explain the
different number of regions detected purely in terms of a differing
spatial resolution.

Based only on broad-band optical imaging data, Loose \& Thuan (1986)
undertook a morphological classification of BCD galaxies. They defined
four different groups:
\begin{itemize}
\item {\bf iE BCDs}. These are galaxies with an outer diffuse halo with
elliptical isophotes and inner irregular isophotes due to the presence of multiple star-forming regions and star clusters.
\item {\bf nE BCDs}. These objects show an elliptical halo and a
clearly defined nucleus.
\item {\bf iI BCDs}. These galaxies have an irregular outer halo and an off-center nucleus. Within this class Loose \& Thuan (1986) also defined
\begin{itemize}
\item[] {\bf iI,C BCDs} -- Cometary morphology.
\item[] {\bf iI,M BCDs} -- Apparent mergers.
\end{itemize}
\item {\bf I0 BCDs}. These are galaxies not showing a diffuse extended component in the broad-band images.
\end{itemize}

The classification of a galaxy as nE or iE BCD is been mainly
determined by the shape of the inner isophotes: regular in the case of
nE, and irregular for the iE BCDs. However, the greyscale maps
presented in Figure~\ref{figure3} show that, in some cases, despite
the continuum isophotes in the inner region being regular, the H$\alpha$
images may show a complex structure composed of several individual HII
regions. An example of this behavior is seen in Mrk~409, where the
inner $R$-band isophotes are regular (which would lead one to classify
this galaxy as a nE BCD), but the H$\alpha$ emission shows the
presence of a ring composed of at least 5 HII regions (see
Figure~\ref{figure3}). In cases like that of Mrk~409 we have made use
of the information provided by the H$\alpha$ images, and the galaxies
have been reclassified as iE BCDs. The presence of a ring of
star-forming regions around the nucleus of the galaxy has been
observed in a total of 4 objects in our sample including Mrk~409,
namely Mrk~86 (see Gil de Paz et al$.$ 2000b,c; 2002), Mrk~400,
Mrk~409, and VCC~0655. The radius of this ring ranges from 0.5 to
1.0\,kpc. In these cases we have added a letter `r' to the
morphological type of the objects.

The results of the morphological classification are given in
Table~\ref{table5}. For the objects with detected H$\alpha$ emission
27 out of 111 (24\%) were classified as nE BCDs, 37 (33\%) as iE, and
39 (35\%) as iI. For the 39 iI BCDs, 11 are ``cometary'' (iI,C) and 10
are ``mergers'' (iI,M). In only 8 galaxies did we not detect any
underlying/extended $R$-band continuum light (that would be otherwise
be associated with an old stellar population) and, consequently, they
were classified as i0 BCDs. Within this group are the objects with the
lowest metallicity in our sample (I~Zw~18, Tol~65, UCM~1612+1308). For
comparison, within their sample of 28 BCDs, Cair\'{o}s et al$.$
(2001a,b) found 21\% of nE, 39\% of iE, and 32\% of iI BCDs (7\% and
14\% of the total were iI,C and iI,M BCDs, respectively).

It is worth mentioning that, although the fraction of BCDs with
elliptical envelopes is $<$60 per cent, the four galaxies where a ring
of star-forming regions was found are all iE BCDs. This fact suggests that
the shaping of these rings requires a highly symmetric distribution
of the galaxy stellar and total mass.

\section{FITS Images}
\label{images}

The FITS copies of the images in this Atlas are publicly available
individually through the NED image server at {\tt
http://nedwww.ipac.caltech.edu/forms/images.html} and as a collection
through a dedicated web page at {\tt
http://nedwww.ipac.caltech.edu/level5/Sept02/Palco\_BCD/frames.html}. The
FITS headers of these images include, in a homogeneous and
self-explanatory way, all the information regarding the image WCS
solution and flux calibration along with the date of observation,
telescope, filter, exposure time, etc (see also Table~\ref{table3}). A
CDROM with the complete set of FITS images may also be requested from
the first author.

\section{Integrated Photometry}
\label{analysis}

In order to obtain integrated photometry of the galaxies in our sample
we edited the final, flux-calibrated images of the Atlas. Field stars
and background galaxies falling near the position of our galaxies were
removed using the IRAF task {\sc credit} within the CRUTIL package. We
did this for all the $B$ and $R$-band images. Emission from field
stars and from background galaxies in the H$\alpha$ images was removed at
the time of the H$\alpha$ continuum subtraction.

The criteria for identifying a region as belonging to the galaxy
(or not) were those used by Gil de Paz et al$.$ (2000b). Briefly,
these criteria are based (1) on the size and ellipticity of the
emitting region compared to the image PSF, (2) the presence of
H$\alpha$ emission associated with the region, and (3) the distance of
the region from the galaxy center (see Gil de Paz et al$.$ 2000b for
more details).

Once all the images were edited we defined two sets of polygonal
apertures. The first set was constructed to include the total
integrated light originating from the galaxy at continuum wavelengths,
and it was identically used to measure both the $B$ and $R$-band
integrated magnitudes. Due to the different spatial distribution and
morphology of the H$\alpha$ emission compared with the continuum, the
integrated H$\alpha$ fluxes were measured using different sets of
polygonal apertures. In both cases, the integrated fluxes were
obtained using the IRAF task {\sc polyphot}. 

The color term required to determine the $B$ and $R$-band magnitudes
of the galaxies was first computed assuming a mean ($B-R$) color of
0.8\,mag and using the color coefficients given in
Table~\ref{table4}. However, the integrated colors derived were in
some cases significantly different from this average value. The final
magnitudes and colors were then iteratively computed using the limit
of the following sequence as the best ($B-R$) color for the galaxy,
\begin{equation}
(B-R)_{i+1}=(B-R)_i+(k_{B,B-R}-k_{R,B-R})\times\left((B-R)_i-(B-R)_{i-1}\right)
\label{sequence}
\end{equation}
where $k_{B,B-R}$ and $k_{R,B-R}$ are the color coefficients for the
$B$ and $R$ bands, respectively. This sequence ranges from $i$=1 to
$n$, where $(B-R)_0$=0.8\,mag and $(B-R)_1$ is the integrated color
initially measured on the images. Convergence ($\Delta$($B-R$)$\ll$0.01)
occurred after a few ($n$$\sim$5-10) iterations. The final magnitudes
and colors are given in Table~\ref{table5}. The errors shown in this
table were obtained by combining the photometry errors given by the
task {\sc polyphot} with those associated with the calibration of the
images. Fluxes, magnitudes and colors shown are corrected for Galactic
extinction (using the values given in Table~\ref{table1} and the
Galactic extinction law of Cardelli, Clayton, \& Mathis 1989), but they are not
corrected for internal extinction. H$\alpha$ fluxes given in
Table~\ref{table5} are also corrected for underlying stellar
absorption adopting an equivalent width of $-$3\,\AA\
(Gonz\'{a}lez-Delgado, Leitherer, \& Heckman 1999).

We have compared our integrated magnitudes measured with the
asymptotic values given by Doublier et al$.$ (1997, 1999) and
Cair\'{o}s et al$.$ (2001b) and the $B_T$ and $R_T$ magnitudes in the
RC3 and ESO-LV catalogs ($B_T$, de Vaucouleurs et al$.$ 1991; $B_T$,
$R_T$; Lauberts \& Valentijn 1989). The mean difference between the
$B$-band magnitudes given by Cair\'{o}s et al$.$ (2001b) and ours for
a total of 14 galaxies in common is $+$0.06\,mag with an rms of
$\pm$0.22\,mag (our magnitudes are marginally brighter). The
comparison between the total $B$-band magnitudes in the RC3 catalogue
and our observed magnitudes gave a difference of $-$0.04$\pm$0.19\,mag
(44 galaxies in common). The largest $B$-band difference
($+$0.36$\pm$0.34\,mag) is obtained when comparing with the results
for 23 galaxies in common with Doublier et al$.$ (1997, 1999). For the
$R$-band data the differences are $-$0.22$\pm$0.20\,mag (11 galaxies)
and $-$0.15$\pm$0.36\,mag (26 galaxies) with respect to the Cair\'{o}s
et al$.$ (2001b) and Doublier et al$.$ (1997, 1999) samples,
respectively. This systematic difference is reduced to
$-$0.05$\pm$0.16\,mag (9 galaxies) when comparison is made with the
$R$-band magnitudes in the ESO-LV catalogue. Note that the comparison
with the RC3 and ESO-LV catalogs was done using observed magnitudes
(i.e$.$ not corrected for Galactic extinction). Finally, we compared
the aperture-photometry data of Cair\'{o}s et al$.$ (2001a) with our
results. This yields differences of $+$0.16$\pm$0.18\,mag (14
galaxies) and $-$0.10$\pm$0.23\,mag (11 galaxies) in the $B$ and $R$
bands, respectively. The existence of these differences is attributed
to (1) intrinsic differences between the extrapolated asymptotic
magnitudes and our aperture-photometry data, (2) the different
galactic-extinction maps used (Burstein \& Heiles 1984 or Schelegel et
al$.$ 1998), and (3) the different methods adopted for the removal of
field stars and background galaxies falling near the position of the
galaxies under study.

Absolute magnitudes and H$\alpha$ luminosities were derived using the
distances given in Table~\ref{table1}. We computed the equivalent
widths of H$\alpha$ by dividing the H$\alpha$ flux by the flux (per
unit wavelength) in the $R$-band after taking into account the added
contribution of H$\alpha$ to the observed $R$-band magnitude itself
(see Appendix~\ref{appendixA} for more details).

In Figures~\ref{figure4}a \& \ref{figure4}b we show the frequency
histograms in ($B-R$) color and $B$-band absolute magnitude. Average
color and absolute magnitude of the galaxies in our sample are
($B-R$)=0.7$\pm$0.3\,mag and M$_B$=$-$16.1$\pm$1.4\,mag. The average
H$\alpha$ luminosity is log(L$_{\mathrm{H}\alpha}$)=40.0$\pm$0.6
(erg\,s$^{-1}$). In Panels~\ref{figure4}c, \ref{figure4}d,
\ref{figure4}e, and \ref{figure4}f we have plotted, respectively, ($B-R$)
vs$.$ M$_B$, EW(H$\alpha$) vs$.$ M$_B$, L$_{\mathrm{H}\alpha}$ vs$.$
M$_B$, and EW(H$\alpha$) vs$.$ ($B-R$), using different symbols for
each morphological type (dots, nE; filled-stars, iE; open-squares, iI;
open-circles, i0).

Figure~\ref{figure4}c shows that fainter BCD galaxies tend to have
bluer colors. Also the galaxies classified as iI and i0 BCD show, on
average, bluer colors than those in the nE and iE classes. This same
difference is also present in the case of the EW(H$\alpha$) (see
Figure~\ref{figure4}d), where iI and i0 BCDs show significantly larger
equivalent widths. The average colors and equivalent widths of the nE
and iE BCDs are ($B-V$)=0.8\,mag and EW(H$\alpha$)=90\,\AA,
respectively, while for the iI and i0 BCDs these values are
($B-V$)=0.5\,mag and EW(H$\alpha$)=200\,\AA. If we consider only the
galaxies classified as iI BCDs the average colors and EW(H$\alpha$)
values derived are 0.6\,mag and 150\,\AA, respectively. Moreover,
Figure~\ref{figure4}c also shows that for ($B-R$)$<$0.5 about 27\% (6
over 22) of the galaxies are nE/iE types while for ($B-R$)$>$0.95 this
percentage goes up to 95\% (18/19). With regard to the equivalent
width of H$\alpha$, Figure~\ref{figure4}d indicates that for
log\,EW(H$\alpha$)$>$2.4 about 23\% (3/13) are nE/iE BCDs while for
log\,EW(H$\alpha$)$<$1.2 the percentage is 100\% (14/14). The lack of
objects showing both low continuum and H$\alpha$ luminosity
(lower-right corner of Figure~\ref{figure4}d) is mainly due to the
selection effects associated with the objective-prism surveys
searching for emission-line galaxies from whose many galaxies in our
sample were selected. In these surveys the probability of detection is
mainly driven by the emission-line flux and its contrast against the
continuum (Salzer 1989). Therefore, objects with low luminosity will
be detected only if the contrast between the line and the continuum
is very strong, in other words, if the equivalent width is large
($>$20\AA\ typically).

The dotted-line in Figure~\ref{figure4}e represents the model
predictions for a composite stellar population formed by a 3.5-Myr-old
burst with Z$_{\odot}$/5 metallicity and 1\% burst strength in mass
superimposed on a 9-Gyr-old underlying stellar population with the
same metallicity. The independent effects of a change in the age of
the burst, the internal extinction and total mass of the galaxy are
also shown. These models were extensively described in
Section~\ref{newclass}.

Finally, in Figure~\ref{figure4}f we show the distribution of our
galaxies in the EW(H$\alpha$) vs$.$ ($B-R$) color diagram along with
the predictions of the same models for different values of the burst
age and burst strength, ranging from 3.5 to 10\,Myr and from 0.01 to
100\%, respectively. This figure shows that in most of the BCDs in our
sample ($\sim$80\%) the presence of an evolved, underlying stellar
population is required, even if a moderate internal extinction of
E($B-V$)=0.2\,mag is assumed and differences in metallicity between
individual galaxies are taken into account. This value of the color
excess corresponds to the most frequently found value in the
spectroscopic atlas of BCDs of Terlevich et al$.$ (1991). The most
metal-poor objects in our sample, however, do not appear to require an
evolved stellar population to reproduce their ($B-R$) colors and
EW(H$\alpha$). But, the ($B-R$) color is not very sensitive to the
presence of an evolved stellar population when the burst strength is
larger than a few percent, so the existence in these galaxies of such
an evolved population cannot be ruled out by these data. The
combination of optical data and deep near-infrared observations is
crucial in solving this problem (James 1994; Doublier et al$.$ 2001b,
Vanzi et al$.$ 2002; Noeske et al$.$ 2003, submitted).

Figure~\ref{figure4}f also confirms (see above) that there is a clear
difference between the properties of the nE/iE BCDs (filled
symbols) and those of the iI/i0 BCDs (open symbols). BCD galaxies
classified as nE and iE types are significantly redder and show lower
EWs in H$\alpha$ than the iI and i0 BCDs. This is probably due to (1)
a lower dust extinction, (2) higher burst strength, and/or (3) lower
metallicity of the iI and i0 galaxies compared to the nE and iE
BCDs. Some differences in this same sense have been already pointed
out by Noeske et al$.$ (2000) for the case of the iI,C (cometary)
BCDs.

It is worth noting that, despite the number of surveys involved,
different selection criteria, and different physical sizes and
environments where these galaxies were discovered, there are
observational properties that are common to all BCD galaxies within
the same morphological class, although with a significant
dispersion. This suggests that the morphology of these galaxies is
direct testimony to their merging and star-formation histories.

In order to show the wide range of morphologies and physical sizes
spanned by these galaxies, and its relation to their luminosities and
optical colors, we have plotted together the $R$-band and H$\alpha$
maps for 80 of the galaxies in the sample set to a common distance and
using a common surface brightness scale (see Figures~\ref{figure5} and
\ref{figure6}). Due to the dense clustering of galaxies at certain
positions in the color-magnitude diagram (see Figure~\ref{figure4}c)
the representation of the complete sample of 114 objects in
Figures~\ref{figure5} and \ref{figure6} is not possible. The $R$-band
and H$\alpha$ images of the galaxies are shown in boxes of fixed
physical size of 5\,kpc on a side using a gray scale ranging from the
sky level (white) to a surface brightness of 21\,mag/arcsec$^{-2}$ in
$R$ and 1.5$\times$10$^{-15}$\,erg\,s$^{-1}$\,cm$^{-2}$\,arcsec$^{-2}$
in H$\alpha$ (black) (these surface brightness are observed values,
except for the highly obscured objects II~Zw~40 and IC~10 which were
corrected for Galactic extinction). Figure~\ref{figure6} shows that,
as commented on above, the largest H$\alpha$ EWs are found within
those objects showing the lowest luminosities and bluest colors
(bottom-right in this figure). These images also graphically
illustrate that BCD galaxies cover, at least, one order of magnitude
in physical size, from $\sim$0.3\,kpc to $\geq$3\,kpc. A more detailed
analysis of the physical size, structure, and population content of
BCDs will be carried out in subsequent papers.

\section{Conclusions}
\label{conclusions}

Summarizing,

\begin{itemize}

\item $B$, $R$, and H$\alpha$ images for a total of 114 galaxies have
been obtained. According to the new set of criteria proposed to define
the Blue Compact Dwarf class of galaxies all except 9 objects
(II~Zw~33, Mrk~7, Tol~1924$-$416, Mrk~314, II~Zw~33B, UGC~4483,
UCM~0049$-$0045, UCM~1446$+$2312, \& VCC~0001) are classified as
BCDs. This represents an increase of a factor of $\sim$4 with respect
to similar previous studies (Doublier et al$.$ 1997, 1999; Cair\'{o}s
et al$.$ 2001a,b). Indeed, previous studies were mostly based on only
broad-band imaging data.

\item The new set of criteria proposed for defining a galaxy as a BCD
includes a limit in $K$-band luminosity (M$_K$$>$$-$21\,mag), the peak
surface brightness ($\mu_{B,\mathrm{peak}}$$<$22\,mag/arcsec$^2$), and
the color at the peak surface brightness
($\mu_{B,\mathrm{peak}}$$-$$\mu_{R,\mathrm{peak}}$\lsim1).

\item The flux-calibrated and WCS-compliant images of the Atlas are
publicly available through the NASA/IPAC Extragalactic Database (NED)
image service on a object-by-objects basis and through a dedicated web
page within LEVEL5: A Knowledgebase for Extragalactic Astronomy \&
Cosmology ({\tt http://nedwww.ipac.caltech.edu/level5/}).

\item In all but three galaxies we detect H$\alpha$-line
emission. About 73\% of the galaxies show H$\alpha$ emission
distributed in more than one region (39\% in 4 or more
regions). Morphologically, 24\% of the galaxies are classified as nE
BCDs, 33\% as iE, 35\% as iI (10\% cometary BCDs -iI,C- and 9\%
mergers -iI,M-), and 7\% as i0 BCDs (see Loose \& Thuan 1986). Four of
the galaxies (Mrk~86, Mrk~400, Mrk~409, VCC~0655) show a nuclear ring of
star-forming regions with radius 0.5-1\,kpc.

\item Average colors, absolute magnitudes and H$\alpha$ luminosities of
the sample are ($B-R$) = 0.7$\pm$0.3 mag, M$_B$=$-$16.1$\pm$1.4\,mag,
and log(L$_{\mathrm{H}\alpha}$)=40.0$\pm$0.6 (erg\,s$^{-1}$). Galaxies
classified as nE and iE BCDs show, on average, redder colors and lower
H$\alpha$ equivalent widths than those classified as iI and i0
BCDs. This is also true if only the iI-type BCDs are considered.

\item For most of the galaxies ($\sim$80\%) the integrated ($B-R$)
colors and H$\alpha$ equivalent widths require the presence of an
evolved, underlying stellar population in addition to a young
population with burst strength lower than 10\% in mass. The most
metal-poor BCDs (I~Zw~18, Tol~65, UCM~1612+1308) are found in that
20\% of the sample whose properties are compatible with the evolution
of a pure, young burst. They also show the bluest colors and highest
equivalent widths within the sample. However, due to the degraded
sensitivity of the ($B-R$) color to the presence of an underlying
stellar population for burst strengths $\geq$10\%, the use of deep
near-infrared imaging data will be required to further investigate, in
a statistical way, their nature as young galaxies.

\end{itemize}

\acknowledgments We are grateful to the Palomar and Las Campanas
observatories staff for their support and hospitality, and to the
Caltech/Palomar and OCIW/Las Campanas Time Allocation Committees for
the generous allocation of time to this project. AGdP acknowledges
financial support from NASA through a Long Term Space Astrophysics
grant to BFM. AGdP is also partially supported by the CONACYT (Mexico)
grant 36132-E and the Spanish Programa Nacional de Astronom\'{\i}a y
Astrof\'{\i}sica under grant AYA2000-1790. This research has made use
of the NASA/IPAC Extragalactic Database (NED) which is operated by the
Jet Propulsion Laboratory, California Institute of Technology, under
contract with the National Aeronautics and Space Administration. We
would like also to thank K. G. Noeske , P. G. P\'{e}rez-Gonz\'{a}lez,
and C. S\'{a}nchez Contreras for valuable discussions and the
NED/LEVEL5 staff for helping us making this Atlas publicly
available. For technical support on using the images from this atlas,
please write to ned@ipac.caltech.edu or to one of the authors (agpaz,
barry, olga@ipac.caltech.edu). We are grateful to the anonymous
referee for her/his helpful comments and suggestions.

\clearpage

\appendix

\section{Emission-line continuum subtraction using broad-band images}
\label{appendixA}

Some of the expressions derived in this appendix also appear in Waller
(1990). Here we have carried out a more detailed analysis of the
procedures used for flux-calibrating broad- or narrow-band imaging
data. We have also considered the case that the field stars used for
the normalization of the continuum image have absorption features in
their spectra and considered the presence of other emission-lines in
the galaxy spectrum (e.g$.$ the
[NII]$\lambda\lambda$6548,6584\,\AA\AA\ doublet in our H$\alpha$
filters) within the narrow-band filter used.

We consider two images, one taken with a narrow-band filter
($\leq$100\,\AA; NB hereafter) and another taken with a broad-band one
($\geq$1000\,\AA; BB hereafter). If we would be observing an
astronomical object having a line in emission within the wavelength
range of these filters, the total fluxes obtained (in counts after the
sky subtraction) would be
\begin{eqnarray}
f_{\,{\rm NB}} = g_{\,{\rm NB}} \times \left( \int f_{{\rm cont},\lambda}\,{\rm S}_{{\rm NB},\lambda}\,{\rm d}\lambda + \int f_{{\rm line},\lambda}\,{\rm S}_{{\rm NB},\lambda}\,{\rm d}\lambda \right) \label{basic1}\\
f_{\,{\rm BB}} = g_{\,{\rm BB}} \times \left( \int f_{{\rm cont},\lambda}\,{\rm S}_{{\rm BB},\lambda}\,{\rm d}\lambda + \int f_{{\rm line},\lambda}\,{\rm S}_{{\rm BB},\lambda}\,{\rm d}\lambda \right), \label{basic2}
\end{eqnarray}
respectively for the NB and BB filters, where $g_{\,{\rm NB}}$ and
$g_{\,{\rm BB}}$ are the ratios between counts and flux in
erg\,s$^{-1}$\,cm$^{-2}$ (including exposure time, gain, system total
efficiency, and atmospheric extinction), $f_{{\rm line},\lambda}$ and
$f_{{\rm cont},\lambda}$ are the fluxes per unit wavelength (in units
of erg\,s$^{-1}$\,cm$^{-2}$\,\AA$^{-1}$) emitted by the object due to
the line and the continuum, respectively, and ${\rm S}_{{\rm
line},\lambda}$ and ${\rm S}_{{\rm cont},\lambda}$ are the normalized
response functions of the NB and BB (including both the filter and
detector efficiency), respectively.

Assuming that the continuum is approximately flat in the spectral
region of both filters, we can write
\begin{eqnarray}
f_{\,{\rm NB}} \simeq g_{\,{\rm NB}} \times \left( {\rm FWHM}_{{\rm NB}}\,f_{{\rm cont},\lambda}
+ \int f_{{\rm line},\lambda}\,{\rm S}_{{\rm NB},\lambda}\,{\rm d}\lambda \right) \\
f_{\,{\rm BB}} \simeq g_{\,{\rm BB}} \times \left( {\rm FWHM}_{{\rm BB}}\,f_{{\rm cont},\lambda} + \int f_{{\rm line},\lambda}\,{\rm S}_{{\rm BB},\lambda}\,{\rm d}\lambda \right)
\end{eqnarray}
where FWHM is the Full Width at Half Maximum of the corresponding filter.

If we now consider that the width of the emission-line is
significantly narrower that the NB filter (less that 1/10$^{{\rm th}}$
the FWHM$_{{\rm NB}}$) we can simplify these expressions to
\begin{eqnarray}
f_{\,{\rm NB}} \simeq g_{\,{\rm NB}} \times ( {\rm FWHM}_{{\rm NB}}\,f_{{\rm cont},\lambda} + f_{{\rm line}}\,{\rm S}_{{\rm NB},{\rm line}} ) \\
f_{\,{\rm BB}} \simeq g_{\,{\rm BB}} \times ( {\rm FWHM}_{{\rm BB}}\,f_{{\rm cont},\lambda} + f_{{\rm line}}\,{\rm S}_{{\rm BB},{\rm line}} ),
\end{eqnarray}
where $f_{{\rm line}}$ is the total flux of the emission line (in
erg\,s$^{-1}$\,cm$^{-2}$) and ${\rm S}_{{\rm NB},{\rm line}}$ and
${\rm S}_{{\rm BB},{\rm line}}$ are the normalized responses of the
filters at the wavelength of the emission line.

Then, defining
\begin{eqnarray}
\beta = {{{\rm FWHM}_{{\rm NB}}} \over {{\rm FWHM}_{{\rm BB}}}} \\
\gamma = {{g_{\,{\rm NB}}} \over {g_{\,{\rm BB}}}} \\
\delta = {{{\rm S}_{{\rm NB},{\rm line}}} \over {\beta}} - {\rm S}_{{\rm BB},{\rm line}}\\
\epsilon = {{{\rm S}_{{\rm BB},{\rm line}}} \over {\rm S}_{{\rm NB},{\rm line}}}
\end{eqnarray}
we obtain $f_{{\rm line}}$, $f_{{\rm cont},\lambda}$, and the equivalent width of the line (EW) as
\begin{eqnarray}
f_{{\rm line}} = {{1} \over {g_{\,{\rm NB}}\,\beta \delta}} \left(\,f_{\,{\rm NB}} - \beta \gamma\,f_{\,{\rm BB}} \right) = {{1} \over {g_{\,{\rm BB}}\,\beta \gamma \delta}} \left( f_{\,{\rm NB}} - \beta \gamma\,f_{\,{\rm BB}} \right)
\label{eq_final} \\
f_{{\rm cont},\lambda} = {{\gamma} \over {g_{\,{\rm NB}}\,{\rm FWHM}_{{\rm NB}}\, \left( {{1} \over {\beta}} - \epsilon \right)\,}} \left(\,f_{\,{\rm BB}} - {{\epsilon}\over{\gamma}}\,f_{\,{\rm NB}} \right) = {{1} \over {g_{\,{\rm BB}}\,{\rm FWHM}_{{\rm BB}}\, \left( 1 - \epsilon \, \beta \right)}} \left(\,f_{\,{\rm BB}} - {{\epsilon}\over{\gamma}}\,f_{\,{\rm NB}} \right) \label{eq_cfinal}\\
{\rm EW}_{\rm line}= {\rm FWHM}_{{\rm BB}} \ {{\left( 1 - \epsilon \beta \right)}\over{\beta \gamma \delta}} \ {{\left( f_{\,{\rm NB}} - \beta \gamma\,f_{\,{\rm BB}} \right)}\over{\left(\,f_{\,{\rm BB}} - {{\epsilon}\over{\gamma}}\,f_{\,{\rm NB}} \right)}}
\label{eq_ewfinal}
\end{eqnarray}

In order to solve these equations (and obtain the pure emission-line
image of the object) we need to determine the $\beta$, $\gamma$,
$\delta$, and $\epsilon$ parameters. The $\beta$, $\delta$, and
$\epsilon$ parameters can be easily obtained from the normalized
response function of the NB and BB filters given the rest-frame
wavelength of the emission-line considered and the recession velocity
of the object.

With regard to the $\gamma$ parameter two approaches can be
followed. First, we can determine the value of the $\gamma$ parameter
if both the NB and BB images are flux-calibrated just dividing the
corresponding calibration factors. However, this parameter can be also
measured without flux-calibrating any of the two images. In this
sense, we can measure the total counts on both images for a large
number of objects with well-known spectral properties in the
wavelength range of interest. In the case of the line equivalent
width, it is not necessary either to calibrate any of the images since
Equation~\ref{eq_ewfinal} does not depend on $g_{\,{\rm NB}}$ or
$g_{\,{\rm BB}}$. On the other hand, if we are interested in deriving
the emission-line flux (or the continuum flux) of the object at least
one of the images should be flux-calibrated.

Thus, with regard to the derivation of $\gamma$, if we measure the
total counts on both the NB and BB for a number of {\it featureless}
objects (usually field stars) we get
\begin{eqnarray}
f_{\,{\rm NB}} = g_{\,{\rm NB}} \times {\rm FWHM}_{{\rm NB}}\,f_{{\rm
cont},\lambda} \label{featureless}\\ f_{\,{\rm BB}} = g_{\,{\rm BB}} \times {\rm
FWHM}_{{\rm BB}}\,f_{{\rm cont},\lambda},
\label{featureless2}
\end{eqnarray}
which leads to
\begin{equation}
\gamma = {{1} \over {\beta}}\,{{f_{\,{\rm NB}}} \over {f_{\,{\rm BB}}}}
\label{gamma}
\end{equation}
This is the most commonly used method for subtracting the continuum
contribution in narrow-band images. This method is particularly simple
and powerful because the ${f_{{\rm NB}}}/{f_{{\rm BB}}}$ ratio
(=$\beta\times\gamma$) can be used to scale the BB image to the NB one
and obtain a pure emission-line image without any previous knowledge
about the filters response functions.

In addition, in some cases it is not possible to find {\it
featureless} stars usually because the emission-line of interest (or
other lines) are present in absorption (or emission) in their
spectra. In that case the Equation~\ref{gamma} has to be substituted by
\begin{equation}
\gamma = {{1 + \sum_i {{{{\rm EW}_{{\rm line,}i}} \over {{\rm FWHM}_{{\rm BB}}}}}\,{\rm S}_{{\rm BB},{\rm line,}i}} \over {\beta + \sum_i {{{{\rm EW}_{{\rm line,}i}} \over {{\rm FWHM}_{{\rm BB}}}}}\,{\rm S}_{{\rm NB},{\rm line,}i}}}\,{{f_{{\rm NB}}} \over {f_{{\rm BB}}}}
\end{equation}
where the sum is extended to the number of {\it features} present in
the field stars spectra (equivalent widths would be negative in
absorption and positive in emission).

The values determined for $\beta$, $\gamma$ (or the
$\beta\times\gamma$ product) would allow to obtain the
continuum-subtracted image. However, if we intend to calibrate the
resulting image resolving $f_{\mathrm{line}}$ in
Equation~\ref{eq_final} we have to previously derive $g_{\,{\rm NB}}$ or
$g_{\,{\rm BB}}$. The observation of several spectrophotometric standards
stars would allow to obtain the following relation
\begin{equation}
m_{{\rm cont},\nu}={\rm ZP}-k_{{\rm NB}} \sec(z) + 2.5 \log(t_{\rm NB}) - 2.5 \log({f_{{\rm NB}}})
\label{mdef}
\end{equation}
where ZP is the zero point of the calibration, $k_{\lambda,{\rm
line}}$ is the extinction coefficient for the NB filter, $t_{\rm NB}$
is the exposure time of the NB image, and $m_{{\rm cont},\nu}$ is the
monochromatic magnitude (see e.g$.$ Hamuy et al$.$ 1992), which is
related with the flux via
\begin{eqnarray}
m_{{\rm cont},\nu}= - 2.5 \log(f_{{\rm cont},\nu}) - 48.590\\
m_{{\rm cont},\nu}= - 2.5 \log(f_{{\rm cont},\lambda}) - 5 \log(\lambda) - 2.398
\label{mnu}
\end{eqnarray}
where $f_{{\rm cont},\nu}$ is expressed in
erg\,s$^{-1}$\,cm$^{-2}$\,Hz$^{-1}$, $f_{{\rm cont},\lambda}$ in
erg\,s$^{-1}$\,cm$^{-2}$\,\AA$^{-1}$, and $\lambda$ is in \AA.

Thus, once the ZP and $k_{{\rm NB}}$ coefficients are obtained from
the Bouger-line fit to the spectrophotometric standards data we can
obtain the conversion factor $g_{\,{\rm NB}}$ from
Equations~\ref{featureless}, \ref{mdef}, and \ref{mnu} as
\begin{equation}
g_{\,{\rm NB}} = {{f_{{\rm NB}}} \over {f_{{\rm cont},\lambda}}} \  {{1} \over {{\rm FWHM}_{\rm NB}}} = 10^{\,0.4 ({\rm ZP}-k_{{\rm NB}} \sec(z) + 2.398)} \, {\lambda^2\,{t_{\rm NB}} \over {{\rm FWHM}_{\rm NB}}}
\end{equation}
and, then
\begin{equation}
f_{\rm line} = 10^{-0.4 ({\rm ZP}-k_{{\rm NB}} \sec(z) + 2.398)} \, 
{{{\rm FWHM}_{\rm NB}} \over {\lambda^2\,t_{\rm NB}\,\beta\,\delta}} \, \left(\,f_{\,{\rm NB}} - \beta \gamma\,f_{\,{\rm BB}} \right)
\label{ffinal1}
\end{equation}

On the other hand, in the case that only the broad-band image is flux
calibrated we obtain
\begin{equation}
m_{{\rm BB}} = {\rm ZP}-k_{{\rm BB}} \sec(z) + k_C C + 2.5
\log(t_{\rm BB}) - 2.5 \log(f_{\rm BB})
\end{equation}
where $k_{{\rm BB}}$ is the extinction coefficient for the BB filter,
$C$ and $k_{C}$ are the color and color coefficient for a
particular color term, and $t_{\rm BB}$ is the exposure time for the
BB image. Since $m_{{\rm BB}}$ is also expressed as
\begin{equation}
m_{{\rm BB}} = m_{{\rm BB, Vega}} - 2.5 \log \left({{f_{{\rm
cont},\lambda}} \over {f_{{\rm cont},\lambda,{\rm Vega}}}}\right)
\end{equation}
we derive $g_{\,{\rm BB}}$ using
\begin{equation}
g_{\,{\rm BB}} = {{f_{{\rm BB}}} \over {f_{{\rm cont},\lambda}}} \  {{1} \over {{\rm FWHM}_{\rm BB}}} = 10^{\,0.4 ({\rm ZP}-k_{{\rm BB}} \sec(z) + k_C C - m_{{\rm BB, Vega}})} \, {{t_{\rm BB}} \over {f_{{\rm cont},\lambda,{\rm Vega}}\,{\rm FWHM}_{\rm BB}}},
\end{equation}
which finally leads to the flux of the emission line
\begin{equation}
f_{\rm line} = 10^{-0.4 ({\rm ZP}-k_{{\rm BB}} \sec(z) + k_C C -
m_{{\rm BB, Vega}})} \, {{f_{{\rm cont},\lambda,{\rm Vega}}\,{\rm
FWHM}_{\rm BB}} \over {t_{\rm BB}\,\beta\,\gamma\,\delta}} \, \left( f_{\,{\rm NB}} - \beta \gamma\,f_{\,{\rm BB}} \right)
\label{ffinal2}
\end{equation}

In those cases where both photometric and spectrophotometric
calibrators are available the comparison of the $g_{\,{\rm
NB}}$/$g_{\,{\rm BB}}$ ratio with the right side of
Equation~\ref{gamma} should provide an additional test for the
reliability of the spectral response functions assumed for the NB and
BB filters. In our case the differences derived between these two
values were of the order of 5 per cent.

Finally, it is important to take into account the contribution that
other emission lines could have to the fluxes and equivalent widths
derived using Equations~\ref{eq_final}, \ref{ffinal1}, \ref{ffinal2},
and \ref{eq_ewfinal}. Although the terms due to these other lines
(typically the doublet [NII]$\lambda\lambda$6548,6584\,\AA\AA\ for
observations in the light of H$\alpha$) are not included in
Equations~\ref{basic1} and \ref{basic2} their contribution can be
corrected by considering that $f_{{\rm line}}$, as it appears in these
equations, can be defined in a way that 
\begin{equation}
{{\rm S}_{{\rm NB},{\rm line}}}\, f_{{\rm line}} \equiv {{\rm S}_{{\rm
NB},{\rm line}}}\, f'_{{\rm line}} + \sum_j {{\rm S}_{{\rm NB},j}}\,
f'_j
\end{equation}
where $f'_{{\rm line}}$ and $f'_j$ are the corrected fluxes for the
line of interest and those other lines included in the filter,
respectively, and the sum in the index $j$ is extended to all
contaminating lines but not the line of study. If we now consider the
line ratios between the contaminating lines and the line of interest
given by spectroscopy observations,
\begin{equation}
r_j \equiv {{f'_j}\over{f'_{{\rm line}}}}
\end{equation}
we obtain
\begin{equation}
f'_{{\rm line}}= \, {{{\rm S}_{{\rm NB},{\rm line}}}\over{{\rm
S}_{{\rm NB},{\rm line}} + \sum_j {\rm S}_{{\rm NB},j}\,r_j}} \
f_{{\rm line}}.
\end{equation}
This expression and the corresponding correction of the EW$_{{\rm
line}}$ are valid as long as the contribution of the contaminating
lines to the flux within the BB filter is negligible.

\clearpage
\begin{deluxetable}{lcccrcccc}
\tablenum{1}
\tabletypesize{\small}
\tablecaption{Sample\label{table1}}
\tablewidth{0pt}
\tablehead{
\colhead{Object Name} & \colhead{RA$_{2000}$}   & \colhead{DEC$_{2000}$}   &
\colhead{A$_B$} & \colhead{$v_{\mathrm{helio}}$}  & \colhead{$\lambda_{\mathrm{H}\alpha}$} & \colhead{H$\alpha$ filter} & \colhead{Distance} &
\colhead{[NII]/H$\alpha$}\\
  \colhead{} &  \colhead{(h m s)} & \colhead{($^{\circ}$ ' '')} & \colhead{(mag)} & \colhead{(km/s)} & \colhead{(\AA)} & \colhead{} & \colhead{(Mpc)} & \colhead{}\\
 \colhead{(1)} & \colhead{(2)} & \colhead{(3)} & \colhead{(4)} & \colhead{(5)} &\colhead{(6)} &\colhead{(7)} &\colhead{(8)} &\colhead{(9)}
}
\startdata
IC~10$^{(1)}$     & 00 20 17.3 &$+$59 18 14 & 6.16 & -348 & 6555.2 & PO 6563 &  1.0 & 0.045 \\ 
HS~0029+1748      & 00 32 03.1 &$+$18 04 46 & 0.57 & 2188 & 6610.7 & PO 6614 & 33.3 & 0.055: \\ 
Haro~14           & 00 45 46.4 &$-$15 35 49 & 0.09 &  944 & 6583.5 & PO 6584 & 14.0 & 0.093 \\ 
UM~285            & 00 51 58.8 &$-$01 40 18 & 0.24 & 1900 & 6604.4 & PO 6601 & 28.2 & 0.055: \\ 
UCM~0049-0045     & 00 51 59.6 &$-$00 29 12 & 0.17 & 1633 & 6598.5 & PO 6593 & 24.4 & 0.055: \\ 
UM~323            & 01 26 46.6 &$-$00 38 46 & 0.13 & 1923 & 6604.9 & PO 6601 & 28.3 & 0.055: \\ 
Mrk~996           & 01 27 35.5 &$-$06 19 36 & 0.19 & 1622 & 6598.3 & PO 6593 & 23.7 & 0.057 \\ 
UM~133            & 01 44 41.3 &$+$04 53 26 & 0.16 & 1623 & 6598.3 & PO 6593 & 24.1 & 0.055: \\
\enddata
\tablecomments{Columns stand for: (1) Galaxy name. (2) Right Ascension (J2000) in (h m s). (3) Declination (J2000) in ($^{\circ}$ ' ''). (4) $B$-band Galactic extinction in magnitudes (Schelegel et
al$.$ 1998) . (5) Heliocentric radial velocity in km\,s$^{-1}$ (taken from NED). (6) Observed wavelength for H$\alpha$ (\AA). (7) Filter used for the H$\alpha$-line imaging. (8) Distance (Mpc). (9) [NII]$\lambda$6584\,\AA/H$\alpha$ intensity ratio. Values of 0.055: are estimated values only, adopted from the average of 49 other galaxies in the sample.}
\tablerefs{
For non-redshift distances: (1) Hunter (2001); (2) Tosi et al$.$ (2001); (3) Sharina, Karachentsev, \& Tikhonov (1999); (4) Dolphin et al$.$ (2001); (5) Meurer et al$.$ (1996); (6) \"Ostlin (2000); (7) Schulte-Ladbeck et al$.$ (1999); (8) Schulte-Ladbeck et al$.$ (2000); (9) Virgo cluster (Macri et al$.$ 1999); (10) Schulte-Ladbeck et al$.$ (2001); (11) Crone et al$.$ (2000); (12) Drozdovsky et al$.$ (2001).}
\end{deluxetable}

\clearpage
\begin{deluxetable}{lccccccc}
\tablenum{2}
\tabletypesize{\small}
\tablecaption{Observing runs\label{table2}}
\tablewidth{0pt}
\tablehead{
\colhead{Run} & \colhead{Nights} & \colhead{Observatory} & \colhead{Telescope} & \colhead{Detector}  & \colhead{Scale} & \colhead{Gain} & \colhead{RN}\\
\colhead{}  &    \colhead{}  &  \colhead{}  & \colhead{}  & \colhead{}  &\colhead{(''/pix)}  & \colhead{(e$^{-}$/ADU)}   & \colhead{(e$^{-}$)}\\
\colhead{(1)}  &    \colhead{(2)}  &  \colhead{(3)}  & \colhead{(4)}  & \colhead{(5)}  & \colhead{(6)}  &\colhead{(7)}  & \colhead{(8)}}
\startdata
Jun01   & 06-09  &  Palomar      &   60" &    CCD\#13   &   0.378  & 1.6  &   6.3\\
Sep01   & 10-12  &  Palomar      &   60" &    CCD\#13   &   0.378  & 1.6  &   6.3\\
Nov01   & 08-09  &  Palomar      &   60" &    CCD\#13   &   0.378  & 1.6  &   6.3\\
Dec01   & 04-09  &  Palomar      &   60" &    CCD\#13   &   0.378  & 1.6  &   6.3\\
Jan02   & 14-16  &  Palomar      &   60" &    CCD\#13   &   0.378  & 1.6  &   6.3\\
Jan02B  & 20-23  &  Palomar      &   60" &    CCD\#13   &   0.378  & 1.6  &   6.3\\
Feb02   & 08-09  &  Las Campanas &  100" &    Tek5      &   0.260  & 2.0  &   6.6\\
Mar02   & 05-09  &  Las Campanas &  100" &    Tek5      &   0.260  & 3.0  &   7.0\\
Mar02B  & 14-19  &  Palomar      &   60" &    CCD\#13   &   0.378  & 1.6  &   6.3\\
May02   & 13-16  &  Palomar      &   60" &    CCD\#13   &   0.378  & 1.6  &   6.3\\
Jul02   & 01-08  &  Palomar      &   60" &    CCD\#13   &   0.378  & 1.6  &   6.3\\
\enddata
\tablecomments{Columns stand for: (1) Observing run. (2) Nights observed. (3) Observatory. (4) Telescope. (5) Detector. (6) Spatial scale at the detector in arcsec/pixel. (8) Gain of the detector in e$^{-}$/ADU. (9) Readout noise of the detector in e$^{-}$. Note that two different gain settings were used during the Las Campanas observing runs.}
\end{deluxetable}

\clearpage
\begin{deluxetable}{lccccccccccccccc}
\tablenum{3}
\tabletypesize{\footnotesize}
\rotate
\tablecaption{Images\label{table3}}
\tablewidth{0pt}
\tablehead{
\colhead{Object Name} & \colhead{Tel} & \multicolumn{4}{c}{$B$-band} & & \multicolumn{4}{c}{$R$-band} & & \multicolumn{4}{c}{H$\alpha$}\\
\cline{3-6} \cline{8-11} \cline{13-16}
\colhead{} & \colhead{} & \colhead{Date} & \colhead{t$_{\mathrm{exp}}$} & \colhead{X} & \colhead{{\tiny PSF}} & & \colhead{Date} & \colhead{t$_{\mathrm{exp}}$} & \colhead{X} & \colhead{{\tiny PSF}} & & \colhead{Date} & \colhead{t$_{\mathrm{exp}}$} & \colhead{X} & \colhead{{\tiny PSF}}\\
  \colhead{}   &\colhead{}  &   \colhead{}         &     \colhead{(s)}            &  \colhead{}  & \colhead{('')} &    & \colhead{}   & \colhead{(s)} & \colhead{}& \colhead{('')} & & \colhead{} & \colhead{(s)} & \colhead{}& \colhead{('')}\\
\colhead{(1)}  &    \colhead{(2)}  &  \colhead{(3)}  & \colhead{(4)}  & \colhead{(5)}  & \colhead{(6)}  &\colhead{}  &\colhead{(7)}  & \colhead{(8)}   & \colhead{(9)} & \colhead{(10)}  &\colhead{}  &\colhead{(11)}  &\colhead{(12)}  &\colhead{(13)}  &\colhead{(14)}}
\startdata
IC~10            & P60  & 01/20/02       & 1$\times$900  & 1.24 & 2.5 &  & 12/06/01       & 3$\times$900  & 1.12 & 1.8 &  & 12/06/01       & 3$\times$1500 & 1.12 & 1.8  \\ 
HS~0029+1748     & P60  & 12/07/01       & 1$\times$900  & 1.08 & 3.5 &  & 11/08/01       & 3$\times$600  & 1.05 & 1.5 &  & 11/08/01       & 2$\times$1800 & 1.15 & 1.5  \\ 
Haro~14          & P60  & 12/07/01       & 1$\times$900  & 1.55 & 4.3 &  & 09/11/01       & 3$\times$600  & 1.66 & 1.9 &  & 09/11/01       & 3$\times$1800 & 1.53 & 1.8  \\ 
UM~285           & P60  & 07/07/02       & 1$\times$900  & 1.70 & 2.0 &  & 07/05/02       & 3$\times$600  & 1.66 & 1.7 &  & 07/05/02       & 3$\times$1200 & 1.45 & 1.8  \\ 
UCM~0049-0045    & P60  & 01/21/02       & 1$\times$900  & 1.48 & 3.2 &  & 01/14/02       & 1$\times$900  & 1.40 & 3.6 &  & 01/14/02       & 2$\times$1200 & 1.55 & 2.9  \\ 
UM~323           & P60  & 07/07/02       & 1$\times$300  & 1.64 & 2.2 &  & 07/03/02       & 3$\times$600  & 1.75 & 1.9 &  & 07/03/02       & 2$\times$1500 & 1.60 & 2.0  \\ 
Mrk~996          & P60  & 12/04/01       & 1$\times$900  & 1.32 & 3.3 &  & 09/12/01       & 3$\times$600  & 1.57 & 1.6 &  & 09/12/01       & 3$\times$1800 & 1.35 & 1.4  \\ 
UM~133           & P60  & 07/07/02       & 1$\times$900  & 1.73 & 2.3 &  & 07/06/02       & 1$\times$600  & 1.75 & 1.5 &  & 07/01/02       & 3$\times$1200 & 1.85 & 2.6  \\
\enddata
\tablecomments{Columns stand for: (1) Galaxy name. (2) Observatory and
telescope (P60: Palomar Observatory 60-inch; C100: Las Campanas
observatory du Pont 100-inch). (3) Date of observation for the
$B$-band image. (4) Number of frames and exposure time per frame for
the $B$-band in seconds. (5) Mean airmass of the $B$-band exposure. (6)
Seeing (FWHM) of the $B$-band image in arcsec. (7-10) The same as
(3-6) for the $R$-band exposure. (11-14) The same as (3-6) for the
H$\alpha$ exposure.}
\end{deluxetable}

\clearpage 
\begin{deluxetable}{lccccccrr}
\tablenum{4}
\tabletypesize{\footnotesize}
\tablecaption{Calibration results\label{table4}}
\tablewidth{0pt}
\tablehead{
\colhead{Night} & \colhead{Telescope} & \colhead{Band} & \colhead{$C_{\lambda}$} & \colhead{$k_{\lambda}$} & \colhead{$k_{\lambda,B-R}$} & \colhead{rms} & \colhead{N} & \colhead{Notes}\\
\colhead{} & \colhead{} & \colhead{} & \colhead{(mag)} & \colhead{(mag/am)} &  \colhead{} & \colhead{(mag)} & \colhead{} & \colhead{}\\
\colhead{(1)} & \colhead{(2)} & \colhead{(3)} & \colhead{(4)} & \colhead{(5)} &  \colhead{(6)} & \colhead{(7)} & \colhead{(8)} & \colhead{(9)}}
\startdata
12/04/01 & P60 & B & 23.760 & 0.264 & $+$0.035 & 0.03 & 49 &    \\
         & P60 & R & 23.787 & 0.120 & $-$0.051 & 0.01 & 34 &    \\
12/07/01 & P60 & B & 23.769 & 0.250 & $+$0.050 & 0.05 & 11 & (a)\\
         & P60 & R & 23.796 & 0.110 & $-$0.050 & 0.04 & 12 & (a)\\
01/20/02 & P60 & B & 23.717 & 0.225 & $+$0.052 & 0.03 & 25 &    \\
         & P60 & R & 23.762 & 0.104 & $-$0.046 & 0.02 & 22 &    \\
01/21/02 & P60 & B & 23.674 & 0.202 & $+$0.069 & 0.04 & 15 &    \\
         & P60 & R & 23.793 & 0.116 & $-$0.050 & 0.08 & 12 & (b)\\
\enddata
\tablecomments{(a) Both the extinction coefficient and the color term are fixed. (b) The color term is fixed. Columns stand for: (1) Date corresponding to the civil time at the start of the observing night. (2) Observatory and telescope (P60: Palomar Observatory 60-inch; C100: Las Campanas Observatory du Pont 100-inch). (3) Band. (4) Photometric zero point in magnitudes. (5) Extinction coefficient in mag/airmass. (6) $B-R$ color coefficient. (7) RMS of the calibration. (8) Number of standard stars used in the fit. (9) Notes.}
\end{deluxetable}

\clearpage
\begin{deluxetable}{lrrrcccrr}
\tablenum{5}
\tabletypesize{\small}
\tablecaption{Integrated Photometry\label{table5}}
\tablewidth{0pt}
\tablehead{
\colhead{Object Name} & \colhead{B} &  \colhead{R}   &  \colhead{B-R} & 
\colhead{$f_{\mathrm{H}\alpha}$} & \colhead{$M_{B}$} & 
\colhead{{\small log}\,$L_{\mathrm{H}\alpha}$} & \colhead{T} & \colhead{EW$_{\mathrm{H}\alpha}$}\\
\colhead{} &  \colhead{{\tiny (mag)}} &  \colhead{{\tiny (mag)}} & \colhead{{\tiny (mag)}} &  \colhead{{\tiny (10$^{-14}$\,erg/s/cm$^{2}$)}} & \colhead{{\tiny (mag)}} & \colhead{{\tiny (erg/s)}} & \colhead{} & \colhead{{\tiny (\AA)}}\\
\colhead{(1)} & \colhead{(2)} & \colhead{(3)} & \colhead{(4)} & \colhead{(5)} &  \colhead{(6)} & \colhead{(7)} & \colhead{(8)} & \colhead{(9)}}
\startdata
IC~10            &  5.75$\pm$0.03 &  5.45$\pm$0.08 &  0.30$\pm$0.08 &   51000$\pm$3700  & -19.25 & 40.79 &  iI,C &  36\\ 
HS~0029+1748     & 16.74$\pm$0.05 & 15.93$\pm$0.14 &  0.81$\pm$0.15 &    19.1$\pm$2.5   & -15.87 & 40.40 &  nE   & 228\\ 
Haro~14          & 13.65$\pm$0.05 & 12.91$\pm$0.14 &  0.74$\pm$0.15 &      34$\pm$4     & -17.08 & 39.90 &  iE   &  22\\ 
UM~285           & 17.00$\pm$0.03 & 17.03$\pm$0.08 & -0.02$\pm$0.08 &     5.8$\pm$0.4   & -15.25 & 39.74 &  iI,C & 187\\ 
UCM~0049-0045    & 14.76$\pm$0.04 & 13.36$\pm$0.25 &  1.39$\pm$0.25 &      20$\pm$4     & -17.18 & 40.14 &  iE   &  20\\ 
UM~323           & 16.09$\pm$0.04 & 15.24$\pm$0.08 &  0.85$\pm$0.09 &    14.1$\pm$1.2   & -16.17 & 40.13 &  iE   &  83\\ 
Mrk~996          & 15.01$\pm$0.03 & 14.08$\pm$0.14 &  0.93$\pm$0.14 &      54$\pm$7     & -16.87 & 40.56 &  nE   & 109\\ 
UM~133           & 15.41$\pm$0.03 & 14.51$\pm$0.14 &  0.90$\pm$0.14 &      20$\pm$3     & -16.50 & 40.15 &  iI,C &  60\\
\enddata
\tablecomments{Columns stand for: (1) Galaxy name (2) $B$-band
integrated magnitude and its 1-$\sigma$ error. (3) $R$-band
integrated magnitude and its 1-$\sigma$ error. (4) ($B-R$) integrated color its 
1-$\sigma$ error. (5) Integrated H$\alpha$ flux and error
in units of 10$^{-14}$\,erg\,s$^{-1}$\,cm$^{-2}$ (corrected for underlying 
stellar absorption). (6) $B$-band absolute magnitude. (7) H$\alpha$
luminosity in erg\,s$^{-1}$. (8) Morphological type according to the
classification of Loose \& Thuan (1986). (9) Equivalent width of
H$\alpha$ in \AA. Magnitudes, colors, fluxes, and luminosities are corrected 
for Galactic extinction but not for internal extinction.}
\end{deluxetable}

\clearpage
\begin{figure}
\figurenum{1}
\epsscale{1.00}
\caption{ {\bf a)} Frequency histogram of the ($B-R$) color at the
peak of the surface-brightness profile for our sample of BCD galaxies
and a sample of dIrr (Parodi et al$.$ 2002) and dE (Jerjen et al$.$
2000) galaxies. An average galactic-extinction correction of
A$_B$=0.1\,mag has been applied to the reference samples data. {\bf
b)} The same for the peak surface brightness. The values obtained for
a sample of dIrr galaxies from Patterson \& Thuan (1996) are also
shown. {\bf c)} ($B-K$)-($B-R$) color-color diagram for the 21
galaxies in our Atlas with 2MASS $K$-band magnitudes available. The
predictions of evolutionary synthesis models are shown (see text for
details). The thick, solid-line represents the best fit to the
data. {\bf d)} Frequency histogram of derived absolute $K$-band
magnitudes obtained applying the relation between the ($B-K$) and
($B-R$) colors shown in Panel {\bf c}.
\label{figure1}}
\end{figure}

\begin{figure}
\figurenum{2}
\epsscale{0.65}
\caption{{\bf a)} Spectral response function of the $B$ and $R$
broad-band filters used at the Palomar Observatory 60-inch telescope
(solid-thin line) and at the Las Campanas Observatory du Pont 100-inch
telescope (solid-thick line) convolved with the quantum efficiency of
the detector. The response function of the filters used by Landolt
(1992a; dotted-line) to construct his list of secondary standards and
those originally used to define the Johnson-Cousins system
(dashed-line; $B$-band: Azusienis \& Straizys 1969; $R$-band: Bessell
1990). {\bf b)} Spectral response function of the narrow-band
H$\alpha$ filters used in the Las Campanas Observatory 100-inch
telescope (from left to right: LC 6570, LC 6600, LC 6630). {\bf c)}
The same as {\bf b)} for the Palomar Observatory 60-inch telescope
(from left to right: PO 6563, PO 6570, PO 6584, PO 6593, PO 6601, PO
6614, PO 6624, PO 6640). Vertical marks indicate the range in
wavelength covered by the redshifted H$\alpha$ line for the galaxies
in our sample.\label{figure2}}
\end{figure}

\begin{figure}
\figurenum{3}
\caption{$R$ (left) and continuum-subtracted H$\alpha$ (right) images of the galaxies in the sample. The compass is 0.5\,kpc in size in each image. A blow-up image of the galaxy nuclear region is also shown.\label{figure3}}
\end{figure}

\begin{figure}
\figurenum{4}
\caption{{\scriptsize {\bf a)} Frequency histogram of the ($B-R$)
color. {\bf b)} Frequency histogram of the absolute magnitude,
M$_B$. {\bf c)} Color vs$.$ M$_B$ diagram. Different symbols are used
for nE, iE, iI, and i0 BCDs (see legend in Panel f). The effect of a
color excess of E($B-V$)=0.2\,mag is shown by arrows. {\bf d)}
EW(H$\alpha$) vs$.$ M$_B$ diagram. {\bf e)} H$\alpha$ luminosity vs$.$
M$_B$ diagram. Dotted-line shows the properties of a composite stellar
population formed by a 3.5-Myr-old burst with Z$_{\odot}$/5
metallicity and 1\% burst strength in mass overimposed on a 9-Gyr-old
underlying stellar population with the same metallicity. The effects
of a change in the age of the burst, the internal extinction and total
mass of the galaxy are also shown. {\bf f)} EW(H$\alpha$) vs$.$
($B-R$) color diagram. The predictions of the models for the stellar
population described above are shown.  Solid lines represent models of
constant burst strength and age between 3.5 and 10\,Myr. Dotted lines
are models with constant age but different burst
strength.}\label{figure4}}
\end{figure}

\begin{figure}
\figurenum{5}
\caption{$R$-band images of 80 of the galaxies in the sample. All panels
are 5\,kpc in size. The grayscale ranges between the value of the sky
(white) and that corresponding to a surface brightness of
21\,mag\,arcsec$^{-2}$ (black). The name of the galaxy is shown at the
bottom-left corner of each panel.\label{figure5}}
\end{figure}

\begin{figure}
\figurenum{6}
\caption{Continuum-subtracted H$\alpha$ images of 80 of the galaxies in
the sample. All panels are 5\,kpc in size. The grayscale ranges
between the value of the sky (white) and that corresponding to a
surface brightness of
1.5$\times$10$^{-15}$\,erg\,s$^{-1}$\,cm$^{-2}$\,arcsec$^{-2}$ (black). The name of the galaxy
is shown at the bottom-left corner of each panel.\label{figure6}}
\end{figure}

\end{document}